# The mechanocaloric potential of spin crossover compounds

Karl G. Sandeman[1,2]

[1]*Department of Physics, Brooklyn College of The City University of New York, 2900 Bedford Avenue, Brooklyn, NY 11210, USA*
[2]*Physics Program, The Graduate Center, CUNY, New York, NY 10016, USA*

**Abstract**
We present a first evaluation of the potential for spin crossover (SCO) compounds to be considered as a new class of giant mechanocaloric effect material. From literature data on the variation of the spin crossover temperature with pressure, we estimate the maximum available adiabatic temperature change for several compounds and the relatively low pressures that may be required to observe these effects.

**Introduction**
Ferroic cooling is a technology in which a solid refrigerant is driven between two states by an applied field.[1] The current growth of research into candidate ferroic refrigerants at room temperature has been fuelled by two factors: firstly the promise of device efficiencies that will be greater than those found in gas-based cooling; and secondly by environmental concerns over the global warming potential of current hydrofluorocarbon-based refrigerants.[2]

From a generalised Gibbs Free Energy it is clear that the materials with the greatest cooling potential are those in which there is a strong variation of order parameter with temperature, and in which the coupling between that order parameter and a conjugate field is strong. If we include all relevant order parameters (such as magnetisation, *M*, electrical polarisaton, *P*, and elastic strain $\epsilon$ and volume *V*):

$$dG = -SdT + \sum_i X_i dY_i$$

$$= -SdT - MdH + Vdp - \epsilon d\sigma + PdE$$

then in all cases where continuum thermodynamics can be applied, the adiabatic application of a field $Y_i$ results in a temperature change $\Delta T_{ad}$ - a so-called caloric effect - that is given by

$$\Delta T_{ad} = -\int_0^{Y_{applied}} \frac{T}{C(Y_i,T)} \left(\frac{\partial X_i}{\partial T}\right) dY_i \ .$$

The isothermal application of a field $Y_i$ that ranges in value from zero to an applied value $Y_{applied}$ will result in an entropy change, $\Delta S$, that is given by:

$$\Delta S(T, Y_{max}) = \int_0^{Y_{applied}} \left(\frac{\partial X_i}{\partial T}\right) dY_i \ .$$

There also exists an energetic sum rule which provides a ceiling to the combination of $\Delta T_{ad}$ and $\Delta S$, depending on the applied field, $Y_{applied}$ and the maximum value of the order parameter, $X_{sat}$:

$$\int_0^\infty \Delta S(T, Y_{applied})\, dT = X_{sat} Y_{applied}$$

Research sub-fields pertaining to particular combinations of $X_i$ and $Y_i$ are now well established and include ferroic cooling effects driven by a magnetic field (magnetocaloric effect, MCE), by an electric field (electrocaloric, ECE), by uniaxial stress (elastocaloric, eCE) or by pressure (barocaloric, BCE). In addition, the term "mechanocaloric", which has previously been used in the field of thermoelastic polymers[3] has recently been revived to group together caloric materials research involving either applied uniaxial or volumetric stress.[4] Of these sub-fields, magnetic cooling, using the MCE, is closest to commercial deployment, and a number of cooling prototypes have been demonstrated.[5] The majority of magnetocaloric materials are intermetallics or oxides,[6] with some of those compounds exhibiting both magnetocaloric and mechanocaloric effects, since the largest caloric effects have been seen in materials with first order phase transitions, where the order parameter and volume/strain are strongly coupled.[7,8] For the same reason, mechanocaloric effects have also been observed in giant ECE materials.[9]

However, there are relatively few investigations thus far of mechanocaloric effects in materials that are not already associated with the MCE or ECE. One example is $Mn_3GaN$, an antiferromagnetic metallic antiperovskite in which there is a volume change at the first order Néel transition. This transition is sensitive to pressure and so barocaloric effects have been observed.[10] Another is ammonium sulphate, a ferrielectric in which large barocaloric effects have been studied around the first order, ferrielectric transition temperature.[11] In this article we examine the available literature on barocaloric effects in another class of material - spin crossover (SCO) compounds - in order to assess their mechanocaloric potential. To our knowledge these materials have not previously been considered by the ferroic cooling community.

**Evaluating caloric potential**
An interesting point of comparison between the caloric families is that, while the saturation magnetisation of many $d$-metal magnetocaloric alloys is limited by the quantum mechanical size of magnetic moments to around 120 $Am^2kg^{-1}$, the quantities $\Delta P$ and $\Delta V$ at a phase transition may be expected to show greater variations. Electrical polarisation is associated with the separation of charge centres and therefore has a great potential for enhancement. Meanwhile, volume changes can vary from fractions of a percent to 20 %.

Perhaps as a result, there is not yet an established definition of the word "giant" in connection with all possible caloric effects, but the term has been applied to magnetocaloric materials in which an adiabatic temperature change, $\Delta T_{ad}$ of 2-3 Kelvin is seen in a 2 Tesla field. It has also been applied to materials in which the latent heat, $T\Delta S$, associated with a first order magnetic phase transition is of the order of 20-30 $MJm^{-3}$ or, equivalently, $\Delta S$ values in the range 0.07-0.1 $JK^{-1}cm^{-3}$ near room temperature. By definition, there is an entropy change at a magnetic phase transition, even without an applied field and so the latter is in some ways an easier condition to satisfy. The observed $\Delta T_{ad}$ depends principally on either the heat capacity of the material or, in the case of some first order compounds, on the variation of transition temperature in the applied field ($\partial T_t/\partial Y_i$). The value of $\partial T_t/\partial Y_i$ can be used to examine how close a ferroic material is to having the maximum possible adiabatic temperature change for a given change of magnetisation, volume, or polarisation (as appropriate), and large values of $\partial T_t/\partial Y_i$ are required in order to avoid the deleterious effects of thermal hysteresis on cycling the applied field.[12] (Concerns about hysteresis loss upon cycling mean that those materials close to (tri)criticality are currently favoured.)

It has been shown that the value of $\partial T_t/\partial Y_i$ can be used a tool to search for large $\Delta T_{ad}$ and $\Delta S$ and hence large relative cooling powers (RCP)s associated with field-driven first order transitions.[6,13] In the simplest approximation, the adiabatic temperature change may be simplified to $\Delta T_{ad}(T, Y_{applied}) = (\partial T_t/\partial Y_i)Y_{applied}$ for values of $(\partial T_t/\partial Y_i)$ that are below the value at which $\Delta T_{ad}(T, Y_{applied})$ is maximised. Beyond the value of $(\partial T_t/\partial Y_i)$ that maximises $\Delta T_{ad}$, $\Delta T_{ad}(T, Y_{applied})$ is instead related to the isothermal entropy change via the heat capacity, $C_p$ through $\Delta T_{ad} = -T\Delta S/C_p$.[6] At the point at which the $\Delta T_{ad}$ value is maximal, either expression may be used. However, it is therefore important to know which régime of $(\partial T_t/\partial Y_i)$ a given material occupies, when predicting $\Delta T_{ad}$ values. In addition to evaluating the regime in which SCO compounds typically sit (with regard applied pressure and its effect on the spin crossover temperature $T_{SCO}$), we will show below that the large values of $\partial T_{SCO}/\partial p$ they frequently possess present an opportunity for low pressure barocaloric effects.

**Spin crossover compounds as caloric materials**
Spin crossover occurs in compounds where the crystal field splitting of *d*-orbitals associated with a magnetic moment is of the order of $k_B T$. The effect was discovered by Cambi[14] and is typically observed in octahedrally coordinated complexes of Fe in, for example, $d^5$ or $d^6$ electronic configurations.[15] As a result, a temperature-induced change of state from a low spin (LS) state that breaks Hund's first rule to a high spin (HS) state can occur at temperature $T_{SCO}$. This change of state has been suggested as a tool for applications including switching and sensing.[16] The low temperature state breaks Hund's rules. Crucially for caloric applications, the change of state from LS to HS can be either continuous or first order,[17] and it can occur at temperatures up to and beyond room temperature.[18]

| Compound | $T_{SCO}$ (K) | $\|\Delta S\|$ (J/K mol) | $\|\Delta S\|$ (J/K kg) | $\partial T_{SCO}/\partial p$ (K/GPa) | max. $\Delta T_{ad}$ | Pressure required (GPa) | Ref. for $T_{SCO}$, $\Delta S$, $\partial T_{SCO}/\partial p$ |
|---|---|---|---|---|---|---|---|
| [Fe(PM-BiA)$_2$(NCS)$_2$] | 170 | 58 | 84 | 66 | 8.4 | 0.12 | 19 and 20 |
| [Fe(phen)$_2$(NCS)$_2$] | 180 | 49 | 92 | 220 | 8.3 | 0.04 | 21 and 22 |
| {Fe[H$_2$B(pz)$_2$]$_2$(bipy)} | 160 | 48 | 95 | 188 | 7.6 | 0.04 | 23, 24, 25 |
| {Fe(pmd)$_2$ [Cu(CN)$_2$]$_2$ } | 140 | 36 | 80.5 | 380 (avg.) | 5.6 | 0.013 | 26 |
| [Fe(pmea)(NCS)$_2$] | 184 | 60 (calc.) | 44 | 146 | 4 | 0.03 | 27 |

**Table 1:** Caloric potential of several SCO compounds, estimated from literature data using available isothermal entropy data, formula weights and a typical specific heat capacity of 2000 Jkg$^{-1}$K$^{-1}$. We see that the adiabatic temperature change available in modest applied pressures is substantial. The magnitude of the isothermal entropy change, $\Delta S$ is shown.

SCO materials are paramagnets and it is found that the driving field most effective in shifting the transition temperature is that of elastic strain or pressure. (The $T_{SCO}$ is sensitive to a magnetic field, but at rates of around 0.03 K/Tesla.[28]) Table 1 gives examples of known pressure variations of several $T_{SCO}$ in compounds studied in the literature; from these values, upper bounds on the $\Delta T_{ad}$ values may be calculated. The application of pressure will stabilise the LS state, resulting in a negative $\Delta S$ value in the cases shown. Since for typical SCO materials, the magnitude of the entropy change associated with spin crossover, $|\Delta S| = 100$ Jkg$^{-1}$K$^{-1}$ and $C_p = 2000$ Jkg$^{-1}$K$^{-1}$,[29] the value of $(\partial T_{SCO}/\partial p)p_{applied}$ at which we must

switch from calculating $\Delta T_{ad}$ using $(\partial T_{SCO}/\partial p)p_{applied}$ to using $-T\Delta S/C_p$ is around 10 K (at 200 K) or 15 K (at 300 K), due to the high heat capacity, per kg, of these materials. In Table 1 we may use the régime for which $\Delta T_{ad} = -T\Delta S/C_p$, in order to estimate the *maximum* possible $\Delta T_{ad}$ values and the value of the $\partial T_{SCO}/\partial p$ to estimate the pressure required to generate this $\Delta T_{ad}$ value.

We note from the table that the available isothermal entropy change is considerable; in molar units its magnitude is typically in the range 50-80 J/K mol (~100-160 J/K kg or ~0.2-0.3 JK$^{-1}$ cm$^{-3}$ given the normal density of spin crossover compounds). This puts the SCO materials in a comparable range of volumetric entropy change density to that of giant magnetocaloric materials (see above). Furthermore, as shown in the table, the pressure required to obtain the maximum $\Delta T_{ad}$ possible is often relatively low, since the $\partial T_{SCO}/\partial p$ values for each material are higher than in the case of intermetallics. As a result, the pressures that may be used to obtain technologically relevant $\Delta T_{ad}$ values may be an order of magnitude lower than those used in experiments on intermetallics already known to the MCE and BCE community. A related conclusion was drawn by Moya et al. in the case of the pressures required to induce the ferrielectric transition in ammonium sulphate. That material had a comparable heat capacity and entropy change (in JK$^{-1}$kg$^{-1}$). (At applied pressure values $p_{applied}$ that are lower than the pressure required to obtain the maximum $\Delta T_{ad}$ value, the adiabatic temperature change is simply given by $(\partial T_{SCO}/\partial p)p_{applied}$.)

Lastly, we make two further remarks. Firstly, we note that the isothermal transition entropy change in materials such as ammonium sulphate and SCOs is rather more similar to that found in intermetallic barocaloric materials such as shape memory alloys and FeRh when it is expressed in units of JK$^{-1}$m$^{-3}$. The technologically relevant unit in the case of MCE materials is certainly JK$^{-1}$m$^{-3}$ since permanent magnetic fields are the cost centre of a magnetic cooling device.[30] The relevant unit for expressing the entropy change of BCE materials may also turn out to be the same. Secondly, the above analysis does not take into account any energy loss due to transition hysteresis in the cyclic application of an applied pressure. Such losses are an important variable that precludes the application of caloric materials with strongly first order transitions, where the reversible component of an isothermal entropy change or adiabatic temperature change may be significantly reduced.[31] Since the purpose of this manuscript is to highlight the potential of the entire SCO class, however, we here only bring to the reader's attention that transition hysteresis will necessarily have to be tuned. There are very good indications that such tunability is possible, for example in the guest dependence of thermal hysteresis in the SCO transitions of a spin crossover framework (SCOF) compound, [Fe$^{II}$(pz) Ni$^{II}$(CN)$_4$]·xGuest .[18]

The above survey of SCO materials is designed to motivate further work including material optimisation for caloric purposes. We note with regard to material design that *ab initio* calculations of even single molecule SCO compounds are non-trivial, given the high degree of dynamical correlations that can be present.[32] However, one significant advantage of SCO compounds is their tunability; a change of ligand can be used to alter the crystal field and thereby the SCO temperature and the order of the phase transition. This could be a very useful attribute, for example in minimising thermal hysteresis. As in studies of other (in particular magneto-) caloric materials, one may also ask what the origin of the transition entropy is in SCO materials, and how it may be maximised. A large percentage of the latent heat at a first order SCO is due to the entropy associated with the giant change in the moment, from LS to HS, or vice versa. However, the strength of so-called cooperativity, the coupling of the change of spin state to the atomic lattice, can lead to a phonon contribution of similar (or greater) size to the magnetic one.[17] Such complementary use of phonon and magnetic entropy allows further opportunities for material optimisation.

## Conclusions

In summary, from a brief consideration of the literature on spin crossover compounds, we may reasonably expect that these are materials that possess mechanocaloric effects which as large as those seen to date in more conventional intermetallic compounds. Moreover, their low density and bulk modulus may make them an attractive technological prospect. There remains much to be done to fully evaluate the caloric potential of these materials experimentally and theoretically, and to explore an optimise a range of materials with spin crossover transitions around room temperature.

## Acknowledgement
We are grateful to I. Gass, M. Halcrow and C. Kepert for useful discussions and The Royal Society and CUNY-Brooklyn College for financial support.